\documentclass[final,1p,times]{elsarticle}

\usepackage{amssymb}

\usepackage{amsfonts}
\usepackage{amsmath}
\usepackage{graphicx}%

\journal{Journal of Magnetism and Magnetic Materials}

\begin{document}

\begin{frontmatter}



\title{Phase correlations and quasicondensate in a two-dimensional ultracold Fermi gas}

\author[label1,label2]{J. Tempere}
\author[label1]{S. N. Klimin}
\address[label1]{Theory of Quantum and Complex systems, Universiteit Antwerpen,
Universiteitsplein 1, B-2610 Antwerpen, Belgium}
\address[label2]{Lyman Laboratory of Physics, Harvard University, Cambridge, MA 02138, USA}

\begin{abstract}
The interplay between dimensionality, coherence and interaction in superfluid
Fermi gases is analyzed by the phase correlation function of the field of
fermionic pairs. We calculate this phase correlation function for a
two-dimensional superfluid Fermi gas with $s$-wave interactions within the
Gaussian pair fluctuation formalism. The spatial behavior of the correlation
function is shown to exhibit a rapid (exponential) decay at short distances
and a characteristic algebraic decay at large distances, with an exponent
matching that expected from Berezinskii-Kosterlitz-Thouless theory of 2D Bose
superfluids. We conclude that the Gaussian pair fluctuation approximation is
able to capture the physics of quasi long-range order in two-dimensional Fermi gases.
\end{abstract}
\begin{keyword}
Low-dimensional quantum gases \sep quasicondensate \sep Correlations
\PACS 03.75.Ss \sep 03.70.+k \sep 05.70.Fh \sep 03.65.Yz
\end{keyword}
\end{frontmatter}



\section{Introduction}

\label{intro}

The low-temperature physics of interacting quantum gases in reduced dimensions
is a subject of intense research. In particular for Bose gases, the physical
picture of superfluidity and of the superfluid-to-normal transition has been
intensely studied. The Berezinskii-Kosterlitz-Thouless phase transition
between the normal and superfluid phases in a trapped atomic Bose gas in two
dimensions (2D) has been observed by Hadzibabic \emph{et al}.
\cite{Hadzibabic}. A recent series of experimental works focused on the
different phases (superfluid, pseudogap, normal) of 2D Bose gases
\cite{Clade,Hung,Ha,Yefsah}. Fermi gases in 2D have been cooled down into the
pseudogap regime \cite{Feld}, but to date the superfluid-to-normal transition
has not been observed in the 2D case. However, in 2D Fermi superfluids the
interplay between dimensionality, coherence and interaction turns out to be
especially interesting since the coherence originates from the interactions,
which is not the case in Bose gases. In this contribution, we investigate this
interplay through the calculation of the phase correlation functions for the
pair field, where the fermionic pairs can be strongly bound Bose-condensed
molecules (BEC regime) or weakly bound Cooper pairs (BCS regime).

The physics of superfluidity in uniform two-dimensional Fermi gases at nonzero
temperatures is not governed by a true pair condensate, because a true
condensed state is destroyed by fluctuations. Rather, the signature of
superfluidity is the appearance of long-range phase correlations. The concept
of this superfluid state, called a quasicondensate, was developed for Bose
gases by Kagan and Popov \cite{Kagan,Popov}. In a quasicondensate state, the
one-body phase correlation function $F\left(  r\right)  $ decays algebraically
at large distances \cite{Bloch2008},%
\begin{equation}
F\left(  r\right)  \propto r^{-\eta}, \label{corf}%
\end{equation}
where $\eta=1/\left(  n_{s}\lambda_{T}^{2}\right)  $, $n_{s}$ is the
superfluid density, and $\lambda_{T}=\left(  2\pi/T\right)  ^{1/2}$ is the
thermal wavelength. The Berezinskii-Kosterlitz-Thouless (BKT) phase transition
\cite{B,KT} occurs at a critical temperature $T=T_{BKT}$ that corresponds to a
universal value $\eta=1/4$ \cite{Nelson1977}, when the superfluid density
jumps from a finite value to zero. Above $T_{BKT}$, the correlation function
$F\left(  r\right)  $ decays exponentially.

The BKT transition for the ultracold Fermi gases in the BCS-BEC crossover has
been theoretically studied using the long-wavelength approximation for the
effective action of the bosonic pair field
\cite{Babaev,Botelho2006,BKT-PRA2009}. The phase correlation functions for
these bosonic pairs has been considered within the same approximation in Ref.
\cite{Salasnich} revealing the algebraic decay in agreement with BKT theory.
However, within the long-wavelength approximation this algebraic decay of the
correlation function $F\left(  r\right)  $ occurs at all $r$, whereas it is
physically expected only at sufficiently large distances. Therefore a
treatment beyond the long-wavelength approach is necessary in order to
describe phase correlations at large and intermediate distances $r$, and to
estimate a spatial range at which the quasi long-range order appears. For
ultracold Bose gases, correlations were theoretically studied using different
approaches, e. g., a modified many-body $T$-matrix theory \cite{Stoof1,Stoof2}
or Monte Carlo calculations \cite{Kagan2000}. For the ultracold Fermi gases,
to the best of our knowledge, such a treatment beyond the long-wavelength
approach has not yet been performed.

In the present work, the phase correlation functions for the pair field are
obtained using the Gaussian pair fluctuation effective action within the path
integral formalism. This investigation has been inspired by the following
reasons. In the long-wavelength treatment of BKT physics in ultracold Fermi
gases in two dimensions \cite{Babaev,Botelho2006,BKT-PRA2009}, the
long-wavelength hydrodynamic approximation is claimed to be \textquotedblleft
non-perturbative\textquotedblright\ contrary to the Gaussian pair fluctuation
(GPF) approach (Ref. \cite{deMelo1993} and its further developments). In other
words, it was claimed that the Gaussian pair fluctuation approach cannot
describe the quasicondensate. Here we show that this claim is not true. In
recent work \cite{ExtGL}, a systematic long-wavelength expansion of the
effective bosonic action has been developed. This expansion is
non-perturbative with respect to the pair field. On the one hand, when we
assume the fluctuations to be slowly varying and perform the long-wavelength
approximation for the GPF fluctuation action, we arrive at a fluctuation
action in the hydrodynamic limit. On the other hand, when we represent the
pair field as a sum of the uniform saddle-point value and a small fluctuation,
and substitute this trial pair field to the long-wavelength action functional
of Ref. \cite{ExtGL}, we must arrive at the same hydrodynamic limit for the
fluctuation action. However, the coefficients of that action functional do not
change when we assume the fluctuations to be small. We can physically expect
that the aforesaid two limiting transitions (a slow varying pair field and
small fluctuations) commute. In this case, (1) the phase correlation functions
calculated using the GPF action will exhibit an algebraic decay $\propto
r^{-\eta}$ at large distances, (2) the values of $\eta$ will coincide with
those following from the long-wavelength theory. A verification of this
hypothesis would help us to bridge the gap between the GPF and
non-perturbative long-wavelength approaches for the ultracold fermions.

\section{Phase correlation function within the Gaussian pair fluctuation
approach}

\label{theory}

The thermodynamics of the ultracold atomic Fermi gases within the Gaussian
pair fluctuation formalism is completely determined by the partition function
represented through the path integral,%
\begin{equation}
\mathcal{Z}\propto\int\mathcal{D}\left[  \varphi^{\dag},\varphi\right]
e^{-S_{fluct}\left[  \varphi^{\dag},\varphi\right]  } \label{Z}%
\end{equation}
where $S_{fluct}\left[  \varphi^{\dag},\varphi\right]  $ is the quadratic
fluctuation action functional. The fluctuation action was derived for the
ultracold fermions in 3D at $T=T_{c}$ in Ref. \cite{deMelo1993}, below $T_{c}$
in Ref. \cite{deMelo1997}, and for imbalanced fermions in Refs.
\cite{TKD2008a,TKD2008b}. The fluctuation action for an imbalanced Fermi gas
with $s$-wave pairing in two dimensions has been derived in Ref.
\cite{NJP2012}. We use this fluctuation action in the present work:%
\begin{align}
S_{fluct}  &  =\sum_{\mathbf{q},m}\left[  M_{1,1}\left(  q,i\Omega_{m}\right)
\varphi_{\mathbf{q},m}^{\dag}\varphi_{\mathbf{q},m}\right. \nonumber\\
&  \left.  +\frac{1}{2}M_{1,2}\left(  \mathbf{q},i\Omega_{m}\right)  \left(
\varphi_{\mathbf{q},m}^{\dag}\varphi_{-\mathbf{q},-m}^{\dag}+\varphi
_{\mathbf{q},m}\varphi_{-\mathbf{q},-m}\right)  \right]  , \label{sfl}%
\end{align}
where $M_{j,k}\left(  q,i\Omega_{m}\right)  $ are the matrix elements of the
inverse pair fluctuation propagator $\mathbb{M}$,
\begin{align}
M_{1,1}\left(  q,i\Omega_{m}\right)   &  =\int\frac{d^{2}\mathbf{k}}{\left(
2\pi\right)  ^{2}}\frac{1}{4E_{\mathbf{k}}E_{\mathbf{k}+\mathbf{q}}}%
\frac{\sinh\left(  \beta E_{\mathbf{k}}\right)  }{\cosh\left(  \beta
E_{\mathbf{k}}\right)  +\cosh\left(  \beta\zeta\right)  }\nonumber\\
&  \times\left(  \frac{\left(  \xi_{\mathbf{k}+\mathbf{q}}+E_{\mathbf{k}%
+\mathbf{q}}\right)  \left(  \xi_{\mathbf{k}}+E_{\mathbf{k}}\right)  }%
{i\Omega_{m}-E_{\mathbf{k}}-E_{\mathbf{k}+\mathbf{q}}}+\frac{\left(
E_{\mathbf{k}+\mathbf{q}}-\xi_{\mathbf{k}+\mathbf{q}}\right)  \left(
\xi_{\mathbf{k}}+E_{\mathbf{k}}\right)  }{i\Omega_{m}-E_{\mathbf{k}%
}+E_{\mathbf{k}+\mathbf{q}}}\right. \nonumber\\
&  \left.  +\frac{\left(  \xi_{\mathbf{k}+\mathbf{q}}+E_{\mathbf{k}%
+\mathbf{q}}\right)  \left(  \xi_{\mathbf{k}}-E_{\mathbf{k}}\right)  }%
{i\Omega_{m}+E_{\mathbf{k}}-E_{\mathbf{k}+\mathbf{q}}}+\frac{\left(
E_{\mathbf{k}+\mathbf{q}}-\xi_{\mathbf{k}+\mathbf{q}}\right)  \left(
\xi_{\mathbf{k}}-E_{\mathbf{k}}\right)  }{i\Omega_{m}+E_{\mathbf{k}%
+\mathbf{q}}+E_{\mathbf{k}}}\right) \nonumber\\
&  -\frac{1}{g},
\end{align}
and%
\begin{align}
M_{1,2}\left(  q,i\Omega_{m}\right)   &  =-\Delta^{2}\int\frac{d^{2}%
\mathbf{k}}{\left(  2\pi\right)  ^{2}}\frac{1}{4E_{\mathbf{k}}E_{\mathbf{k}%
+\mathbf{q}}}\frac{\sinh\left(  \beta E_{\mathbf{k}}\right)  }{\cosh\left(
\beta E_{\mathbf{k}}\right)  +\cosh\left(  \beta\zeta\right)  }\nonumber\\
&  \times\left(  \frac{1}{i\Omega_{m}-E_{\mathbf{k}}-E_{\mathbf{k}+\mathbf{q}%
}}-\frac{1}{i\Omega_{m}-E_{\mathbf{k}}+E_{\mathbf{k}+\mathbf{q}}}\right.
\nonumber\\
&  \left.  +\frac{1}{i\Omega_{m}+E_{\mathbf{k}}-E_{\mathbf{k}+\mathbf{q}}%
}-\frac{1}{i\Omega_{m}+E_{\mathbf{k}}+E_{\mathbf{k}+\mathbf{q}}}\right)  ,
\end{align}
where $\xi_{\mathbf{k}}=\frac{\hbar^{2}k^{2}}{2m}-\mu$ are the fermion
energies counted from the average chemical potential of \textquotedblleft
spin-up\textquotedblright\ and \textquotedblleft spin-down\textquotedblright%
\ fermions $\mu=\left(  \mu_{\uparrow}+\mu_{\downarrow}\right)  /2$.
Furthermore, $E_{\mathbf{k}}=\sqrt{\xi_{\mathbf{k}}^{2}+\Delta^{2}}$ are the
Bogoliubov excitation energies, $\zeta=\left(  \mu_{\uparrow}-\mu_{\downarrow
}\right)  /2$ is a measure of imbalance through the chemical potentials, and
$\Omega_{m}=2\pi m/\beta$ are the bosonic Matsubara frequencies with
$\beta=1/\left(  k_{B}T\right)  $. In the chosen system of units, $\hbar=1$,
the fermion mass $m=1/2$ and the Fermi energy of a free-fermion gas in 2D is
$E_{F}\equiv\pi\hbar^{2}n/m=1$, where $n$ is the total fermion density. The
coupling strength $g$ is renormalized through the binding energy of a
two-particle bound state $E_{b}$, in the same way as in Ref. \cite{Duan}:%
\begin{equation}
\frac{1}{g}=\frac{1}{8\pi}\left(  \ln\frac{E_{b}}{E}+i\pi\right)  -\int
\frac{d^{2}\mathbf{k}}{\left(  2\pi\right)  ^{2}}\frac{1}{2k^{2}-E+i\delta},
\label{renorm}%
\end{equation}
with $\delta$ a positive infinitesimal.

Within the Gaussian pair fluctuation approach
\cite{deMelo1993,deMelo1997,TKD2008a,TKD2008b,NJP2012} the action is expanded
quadratically around a (uniform) saddle-point $\Delta$. Therefore the
amplitude and the phase components of the fluctuation field variables can be
written as follows:%
\begin{align*}
\varphi\left(  \mathbf{r},\tau\right)   &  =a\left(  \mathbf{r},\tau\right)
+i\Delta\cdot\theta\left(  \mathbf{r},\tau\right)  ,\\
\varphi^{\dag}\left(  \mathbf{r},\tau\right)   &  =a\left(  \mathbf{r}%
,\tau\right)  -i\Delta\cdot\theta\left(  \mathbf{r},\tau\right)  .
\end{align*}
The fluctuation coordinates in the coordinate/time representation
$\varphi\left(  \mathbf{r},\tau\right)  $ are expressed through their Fourier
amplitudes entering (\ref{sfl}):%
\begin{align}
\varphi\left(  \mathbf{r},\tau\right)   &  =\frac{1}{L\sqrt{\beta}}%
\sum_{\mathbf{q}}\sum_{m=-\infty}^{\infty}e^{i\mathbf{q\cdot r}-i\Omega
_{m}\tau}\varphi_{\mathbf{q},m},\label{Fourier}\\
\varphi^{\dag}\left(  \mathbf{r},\tau\right)   &  =\frac{1}{L\sqrt{\beta}}%
\sum_{\mathbf{q}}\sum_{m=-\infty}^{\infty}e^{-i\mathbf{q\cdot r}+i\Omega
_{m}\tau}\varphi_{\mathbf{q},m}^{\dag}. \label{Fourier1}%
\end{align}
The Fourier components of the amplitude and phase are expressed through
$\varphi_{\mathbf{q},m}$ as:%
\begin{equation}
a_{\mathbf{q},m}=\frac{\varphi_{\mathbf{q},m}+\varphi_{-\mathbf{q},-m}^{\dag}%
}{2},\quad\theta_{\mathbf{q},m}=\frac{\varphi_{\mathbf{q},m}-\varphi
_{-\mathbf{q},-m}^{\dag}}{2i\Delta}. \label{atheta}%
\end{equation}
Note that although $\theta\left(  \mathbf{r},\tau\right)  $ is an angular
field, we do not take this periodicity into account, as usual in this formalism.

Let us introduce the matrix elements which are even $\left(  e\right)  $ and
odd $\left(  o\right)  $ with respect to $\Omega_{m}$:%
\begin{align}
M_{1,1}^{\left(  e\right)  }\left(  q,i\Omega_{m}\right)   &  =\frac{1}%
{2}\left[  M_{1,1}\left(  q,i\Omega_{m}\right)  +M_{1,1}\left(  q,-i\Omega
_{m}\right)  \right]  ,\nonumber\\
M_{1,1}^{\left(  o\right)  }\left(  q,i\Omega_{m}\right)   &  =\frac{1}%
{2}\left[  M_{1,1}\left(  q,i\Omega_{m}\right)  -M_{1,1}\left(  q,-i\Omega
_{m}\right)  \right]  .
\end{align}
The matrix element $M_{1,2}\left(  q,i\Omega_{m}\right)  $ is even. The action
functional for the Gaussian fluctuations is then rewritten in terms of the
amplitude and phase fluctuations:%
\begin{align}
S_{fluct}  &  =\sum_{\mathbf{q},m}\left\{  \left[  M_{1,1}^{\left(  e\right)
}\left(  q,i\Omega_{m}\right)  +M_{1,2}\left(  q,i\Omega_{m}\right)  \right]
a_{\mathbf{q},m}^{\dag}a_{\mathbf{q},m}\right. \nonumber\\
&  +\left[  M_{1,1}^{\left(  e\right)  }\left(  q,i\Omega_{m}\right)
-M_{1,2}\left(  q,i\Omega_{m}\right)  \right]  \Delta^{2}\theta_{\mathbf{q}%
,m}^{\dag}\theta_{\mathbf{q},m}\nonumber\\
&  \left.  +M_{1,1}^{\left(  o\right)  }\left(  q,i\Omega_{m}\right)
i\Delta\left(  a_{\mathbf{q},m}^{\dag}\theta_{\mathbf{q},m}-\theta
_{\mathbf{q},m}^{\dag}a_{\mathbf{q},m}\right)  \right\}  . \label{S1}%
\end{align}
The analogous amplitude-phase representation was considered for the
fluctuation action of a balanced Fermi gas in 3D below $T_{c},$ Ref.
\cite{deMelo1997}. In general, the amplitude and phase fluctuations are
coupled in (\ref{S1}). They are decoupled only for $\Omega_{m}=0$, because
$M_{1,1}^{\left(  o\right)  }\left(  q,0\right)  =0$.

The fluctuation action is quadratic, so that the correlation functions of the
field variables are calculated in a straightforward way. We need to determine
the amplitude-amplitude, phase-phase and amplitude-phase quadratic correlation
functions. They are expressed through the matrix elements as follows:%
\begin{align}
\left\langle a_{\mathbf{q},m}a_{\mathbf{q}^{\prime},m^{\prime}}^{\dag
}\right\rangle  &  =\delta_{\mathbf{q}^{\prime},\mathbf{q}}\delta_{m^{\prime
},m}\frac{1}{2}\frac{M_{1,1}^{\left(  e\right)  }\left(  q,i\Omega_{m}\right)
-M_{1,2}\left(  q,i\Omega_{m}\right)  }{\det\mathbb{M}\left(  q,i\Omega
_{m}\right)  },\\
\left\langle \theta_{\mathbf{q},m}\theta_{\mathbf{q}^{\prime},m^{\prime}%
}^{\dag}\right\rangle  &  =\delta_{\mathbf{q}^{\prime},\mathbf{q}}%
\delta_{m^{\prime},m}\frac{1}{2\Delta^{2}}\frac{M_{1,1}^{\left(  e\right)
}\left(  q,i\Omega_{m}\right)  +M_{1,2}\left(  q,i\Omega_{m}\right)  }%
{\det\mathbb{M}\left(  q,i\Omega_{m}\right)  },\\
\left\langle a_{\mathbf{q},m}\theta_{\mathbf{q}^{\prime},m^{\prime}}^{\dag
}\right\rangle  &  =-\delta_{\mathbf{q}^{\prime},\mathbf{q}}\delta_{m^{\prime
},m}\frac{i}{2\Delta}\frac{M_{1,1}^{\left(  o\right)  }\left(  q,i\Omega
_{m}\right)  }{\det\mathbb{M}\left(  q,i\Omega_{m}\right)  },\\
\left\langle a_{\mathbf{q},m}^{\dag}\theta_{\mathbf{q}^{\prime},m^{\prime}%
}\right\rangle  &  =\delta_{\mathbf{q}^{\prime},\mathbf{q}}\delta_{m^{\prime
},m}\frac{i}{2\Delta}\frac{M_{1,1}^{\left(  o\right)  }\left(  q,i\Omega
_{m}\right)  }{\det\mathbb{M}\left(  q,i\Omega_{m}\right)  }.
\end{align}

We consider the phase fluctuation correlation functions%
\begin{equation}
F\left(  \mathbf{r}-\mathbf{r}^{\prime},\tau-\tau^{\prime}\right)
\equiv\left\langle e^{i\theta\left(  \mathbf{r},\tau\right)  }e^{-i\theta
\left(  \mathbf{r}^{\prime},\tau^{\prime}\right)  }\right\rangle _{S_{fluct}}.
\label{F}%
\end{equation}
For a Gaussian fluctuation action, Wick's decomposition theorem allows to
express the correlation function (\ref{F}) as%
\begin{equation}
F\left(  \mathbf{r},\tau\right)  =e^{-G\left(  \mathbf{r},\tau\right)  }
\label{CF2}%
\end{equation}
with the quadratic phase correlator,%
\begin{equation}
G\left(  \mathbf{r},\tau\right)  \equiv\frac{1}{2}\left\langle \left[
\theta\left(  \mathbf{r},\tau\right)  -\theta\left(  0,0\right)  \right]
^{2}\right\rangle _{S_{fluct}}. \label{G}%
\end{equation}
The analogous treatment for Bose gases in 2D was performed in Refs.
\cite{Stoof1,Stoof2}, and for a Fermi gas in 2D within the long-wavelength
approximation in Ref. \cite{Salasnich}.

Using the Fourier expansion for the phase and the obtained quadratic
correlators, we arrive at the result:%
\begin{align}
G\left(  r,\tau\right)   &  =\frac{1}{4\pi\Delta^{2}}\int_{0}^{\infty}%
qdq\frac{1}{\beta}\sum_{m=-\infty}^{\infty}\left[  1-J_{0}\left(  qr\right)
e^{-i\Omega_{m}\tau}\right] \nonumber\\
&  \times\frac{M_{1,1}^{\left(  e\right)  }\left(  q,i\Omega_{m}\right)
+M_{1,2}\left(  q,i\Omega_{m}\right)  }{\det\mathbb{M}\left(  q,i\Omega
_{m}\right)  }. \label{G3}%
\end{align}

In order to analyze the instantaneous behavior of the phase, we consider the
case $\tau=0$ and the functions $F\left(  r\right)  \equiv F\left(
r,0\right)  $ and $G\left(  r\right)  \equiv G\left(  r,0\right)  .$ The
bosonic Matsubara summations are performed using contour integrations
analogously to Refs. \cite{deMelo1993,deMelo1997,TKD2008a,TKD2008b,NJP2012}.
The result is:%
\begin{equation}
G\left(  r\right)  =\frac{1}{4\pi\Delta^{2}}\int_{0}^{\infty}qdq\left[
1-J_{0}\left(  qr\right)  \right]  S\left(  q\right)  , \label{G4}%
\end{equation}
with $\delta\rightarrow+0$ and with the spectral function%
\begin{equation}
S\left(  q\right)  =\frac{1}{\pi}\int_{-\infty}^{\infty}d\omega\frac
{1}{1-e^{-\beta\omega}}\operatorname{Im}\left(  \frac{M_{1,1}\left(
q,\omega+i\delta\right)  +M_{1,2}\left(  q,\omega+i\delta\right)  }%
{\det\mathbb{M}\left(  q,\omega+i\delta\right)  }\right)  . \label{S}%
\end{equation}
The integral over $q$ in (\ref{G4}) is free from the long-wavelength
divergence at $q\rightarrow0$ due to the factor $1-J_{0}\left(  qr\right)  $,
but an ultraviolet divergence at $q\rightarrow\infty$ appears, similarly as in
Refs. \cite{Stoof1,Stoof2}. Namely, the spectral function $S\left(  q\right)
$ decays as $q^{-2}$ at $q\rightarrow\infty$. Therefore the integral $\int
_{0}^{\infty}S\left(  q\right)  qdq$ in (\ref{G4}) diverges logarithmically at
the upper bound, while the other integral, $\int_{0}^{\infty}J_{0}\left(
qr\right)  S\left(  q\right)  qdq$, converges.

The divergent integral $\int_{0}^{\infty}S\left(  q\right)  qdq$ does not
depend on the distance $r$. Hence this divergence leads to an infinite factor
which does not depend on $r$ and hence does not influence the decay rate of
the correlation function. Because this factor is one and the same for all $r$,
the spatial behavior of the correlation functions can be analyzed considering
the relative (fractional) correlation function%
\begin{equation}
F^{R}\left(  r,r_{c}\right)  \equiv\frac{F\left(  r\right)  }{F\left(
r_{c}\right)  } \label{Fr}%
\end{equation}
with an arbitrary $r_{c}$. This fractional correlation function is expressed
as%
\begin{equation}
F^{R}\left(  r,r_{c}\right)  =e^{-G^{R}\left(  r,r_{c}\right)  } \label{Fr1}%
\end{equation}
with the quadratic phase correlator,%
\begin{equation}
G^{R}\left(  r,r_{c}\right)  =\frac{1}{4\pi\Delta^{2}}\int_{0}^{\infty
}qdq\left[  J_{0}\left(  qr_{c}\right)  -J_{0}\left(  qr\right)  \right]
S\left(  q\right)  . \label{GR1}%
\end{equation}

As follows immediately from (\ref{Fr}), the fractional correlation function
$F^{R}\left(  r,r_{c}\right)  $ at $r=0$ becomes equal to $1/F\left(
r_{c}\right)  $, so that the aforesaid ultraviolet divergence again appears at
$r=0$. The divergence of $F^{R}\left(  r,r_{c}\right)  $ at $r=0$ is then an
artifact of the regularization used here. This is a common feature with the
correlation function derived in Ref. \cite{Salasnich} where $F\left(
r\right)  \propto r^{-\eta}$ for all $r$. In the present treatment the
correlation function increases at $r\rightarrow0$ logarithmically, i. e., more
slowly than within the long-wavelength approximation.

The physical reason of the ultraviolet divergence discussed above is the
restriction of the treatment to the Gaussian fluctuations about the saddle
point. A convergent integral over the momentum $q$ might be obtained by a
(partial) series summations over higher-order terms in powers of the
fluctuations. A complete regularization of the correlation function including
the point $r=0$ is beyond the scope of the present work.

It can be shown that the long-range behavior of the correlation functions can
be insensitive to the concrete way of regularization of the integral in
$G\left(  r\right)  $. Let us assume that the spectral function $S\left(
q\right)  $ is renormalized, $S\left(  q\right)  \rightarrow S^{reg}\left(
q\right)  $, in such a way that the integral
\begin{equation}
G^{reg}\left(  r\right)  =\frac{1}{4\pi\Delta^{2}}\int_{0}^{\infty}qdq\left[
1-J_{0}\left(  qr\right)  \right]  S^{reg}\left(  q\right)  \label{Greg}%
\end{equation}
converges. We assume also that the renormalized spectral function
$S^{reg}\left(  q\right)  $ decays faster than $S\left(  q\right)  $ at large
$q$ but tends to $S\left(  q\right)  $ at small $q$. At large distances,
$J_{0}\left(  qr\right)  $ is small, and hence we arrive at the logarithmic
small $q$ divergence in (\ref{Greg}) at $r\rightarrow\infty$. Therefore the
small $q$ range is crucial for the increase of $G^{reg}\left(  r\right)  $ at
large $r$. As long as $S^{reg}\left(  q\right)  \rightarrow S\left(  q\right)
$ at small $q$, $G^{reg}\left(  r\right)  $ is not sensitive to a behavior of
$S^{reg}\left(  q\right)  $ at large $q$.

In order to verify this reasoning, we consider a simple alternative method of
regularization for the correlation function $G\left(  r\right)  $ introducing
an ultraviolet cutoff $q_{c}$ for the momentum $q$, so that the regularized
correlation function is%
\begin{equation}
G^{reg}\left(  r,q_{c}\right)  =\frac{1}{4\pi\Delta^{2}}\int_{0}^{q_{c}%
}qdq\left[  1-J_{0}\left(  qr\right)  \right]  S\left(  q\right)  .
\label{GR2}%
\end{equation}
The regularized correlation function
\begin{equation}
F^{reg}\left(  r,q_{c}\right)  =e^{-G^{reg}\left(  r,q_{c}\right)  }
\label{Freg}%
\end{equation}
is convergent for all $r$.

As shown above, the asymptotic behavior or the correlation functions at
sufficiently large $r$ is determined by the spectral function in the small $q$
region. Therefore the cutoff regularization should not influence this
asymptotic behavior. We thus expect that the fractional correlation functions
(\ref{Fr}) and the regularized correlation functions with (\ref{GR2}) decay in
one the same way at large $r$. This conclusion will be numerically verified in
the next section.

\section{Results and discussion}

\label{results}

In this section the spatial profile of the phase correlation function is
discussed for different temperatures and binding energies. It is especially
interesting to compare the decay of the phase correlations described by
formulae (\ref{Fr1}) and (\ref{GR1}) with the algebraic decay for a
quasicondensate with the power index $\eta=1/\left(  n_{s}\lambda_{T}%
^{2}\right)  $ following from the BKT theory.

Within the microscopic BKT theory of the superfluidity for ultracold fermions
in two dimensions, the parameters of the superfluid density $n_{s}\left(
T,\mu,\zeta,\Delta\right)  $ (the chemical potentials $\mu,\zeta$ and the gap
$\Delta$) are determined from a joint solution of the gap equation and
equations normalizing the total fermion density $n$ and the density difference
$\delta n=n_{\uparrow}-n_{\downarrow}$ (the number equations). It should be
noted that the superfluid density entering the long-wavelength action
functionals \cite{Babaev,Botelho2006,BKT-PRA2009} as a prefactor at $\left(
\nabla\theta\right)  ^{2}$ (e. g., formula (22) of Ref. \cite{BKT-PRA2009}),
\begin{align}
n_{s}(T,\mu,\zeta,\Delta)  &  =\frac{1}{4\pi}\int_{0}^{\infty}dk\text{
}k\left\{  1-\frac{\xi_{k}}{E_{k}}\frac{\sinh\left(  \beta E_{\mathbf{k}%
}\right)  }{\cosh\left(  \beta E_{\mathbf{k}}\right)  +\cosh\left(  \beta
\zeta\right)  }\right. \nonumber\\
&  \left.  -k^{2}\beta\frac{\cosh\beta E_{\mathbf{k}}\cosh\left(  \beta
\zeta\right)  +1}{\left[  \cosh\beta E_{\mathbf{k}}+\cosh\left(  \beta
\zeta\right)  \right]  ^{2}}\right\}  . \label{ns}%
\end{align}
is the \emph{mean-field} expression, because the fluctuation correction to the
mean-field action up to quadratic order is already contained in $\left(
\nabla\theta\right)  ^{2}$.

If one accounts for fluctuations through the chemical potentials entering the
superfluid density, $n_{s}$ must be necessarily completed with a fluctuation
contribution as $n_{s}^{\left(  tot\right)  }=n_{s}^{\left(
mean-field\right)  }+n_{s}^{\left(  fluct\right)  }$. However, the fluctuation
contribution $n_{s}^{\left(  fluct\right)  }$ is not present as a prefactor of
a fluctuation field in the quadratic Gaussian action: it can only appear as a
fluctuation field prefactor in the next (quartic) order correction to
$S_{fluct}$. Thus an account of the influence of fluctuations on the
prefactors in the Gaussian fluctuation action would be, strictly speaking,
beyond the quadratic approximation. Hence it is consistent to calculate the
mean-field superfluid density (\ref{ns}) with the parameters determined using
the mean-field number equations. This principle is held in all known works on
the microscopic BKT theory, including Refs.
\cite{Babaev,Botelho2006,BKT-PRA2009}. For the same reason, in order to
adequately compare the decay of the correlation functions derived in Section
\ref{theory} with that following from the microscopic BKT theory, the matrix
elements $M_{j,k}$ must be calculated with the mean-field values of the
chemical potentials.%

\begin{figure}
[h]
\begin{center}
\includegraphics[
height=4.831in,
width=4.2231in
]%
{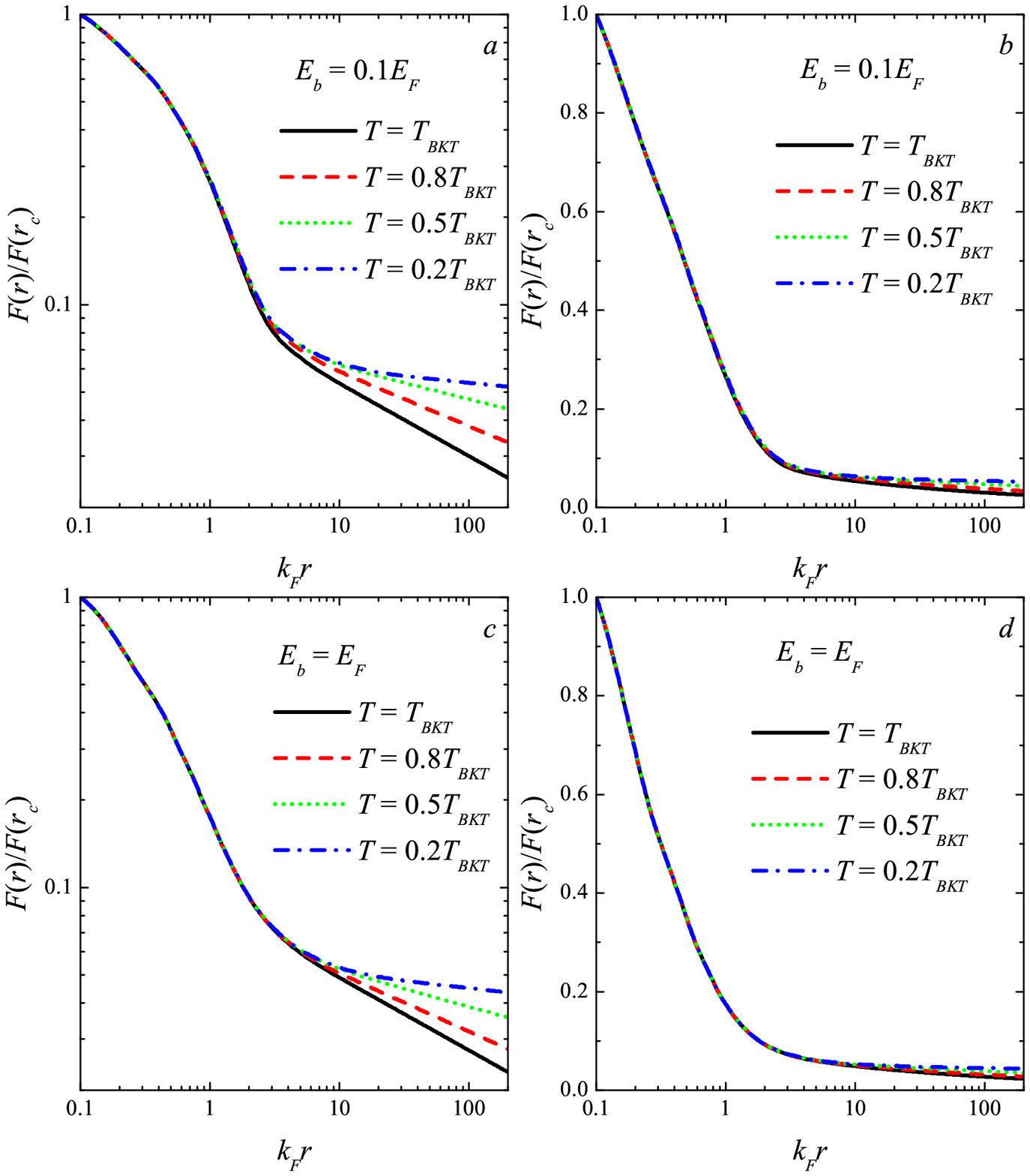}%
\caption{Phase correlation functions for the Fermi gas in 2D with the binding
energy $E_{b}=0.1E_{F}$ (\emph{a}, \emph{b}) and $E_{b}=E_{F}$ (\emph{c}, d).
In panels \emph{a} and \emph{c}, the results are represented on a logarithmic
scale for both $x$ and $y$ axes. In panels \emph{b} and \emph{d}, the
logarithmic scale is used for the $x$ axis only.}%
\label{Fig1}%
\end{center}
\end{figure}

In Fig. \ref{Fig1}, the phase correlation functions are plotted for the Fermi
gas at different temperatures in 2D, for binding energies $E_{b}=0.1$ $E_{F}$
(BCS regime) and $E_{b}=E_{F}$ (BEC-BCS crossover regime). The behavior of the
phase correlation function qualitatively agrees with the result of Ref.
\cite{Kagan2000} for a Bose gas in 2D using the Monte Carlo technique. At
small and intermediate distances, the correlation function decreases rapidly,
obeying closely an exponential decay law. At sufficiently large distances, the
correlation function decreases much slower, and the decay tends to a power law.

In order to characterize the power-law decay, we analyze the parameter
$\alpha$ determined through the logarithmic derivative of the phase
correlation function, defined as%
\begin{equation}
\alpha\left(  r\right)  \equiv-\frac{r}{F\left(  r\right)  }\frac{\partial
F\left(  r\right)  }{\partial r}=r\frac{\partial G\left(  r\right)  }{\partial
r}. \label{alp1}%
\end{equation}
Note that when $F\left(  r\right)  \propto r^{-\eta}$, we obtain
$\alpha\left(  r\right)  =\eta$.

The derivative is determined straightforwardly using formula (\ref{GR1}):%
\begin{align}
\alpha\left(  r\right)   &  =\frac{r}{4\pi\Delta^{2}}\int_{0}^{\infty}%
q^{2}dqJ_{1}\left(  qr\right) \nonumber\\
&  \times\frac{1}{\pi}\int_{-\infty}^{\infty}d\omega\operatorname{Im}\left(
\frac{1}{1-e^{-\beta\left(  \omega+i\delta\right)  }}\frac{M_{1,1}\left(
q,\omega+i\delta\right)  +M_{1,2}\left(  q,\omega+i\delta\right)  }%
{\det\mathbb{M}\left(  q,\omega+i\delta\right)  }\right)  . \label{alp}%
\end{align}
As distinct from the phase correlation function, the integral in (\ref{alp})
is convergent and does not require regularization.

\begin{figure}
[h]
\begin{center}
\includegraphics[
height=4.5455in,
width=3.461in
]%
{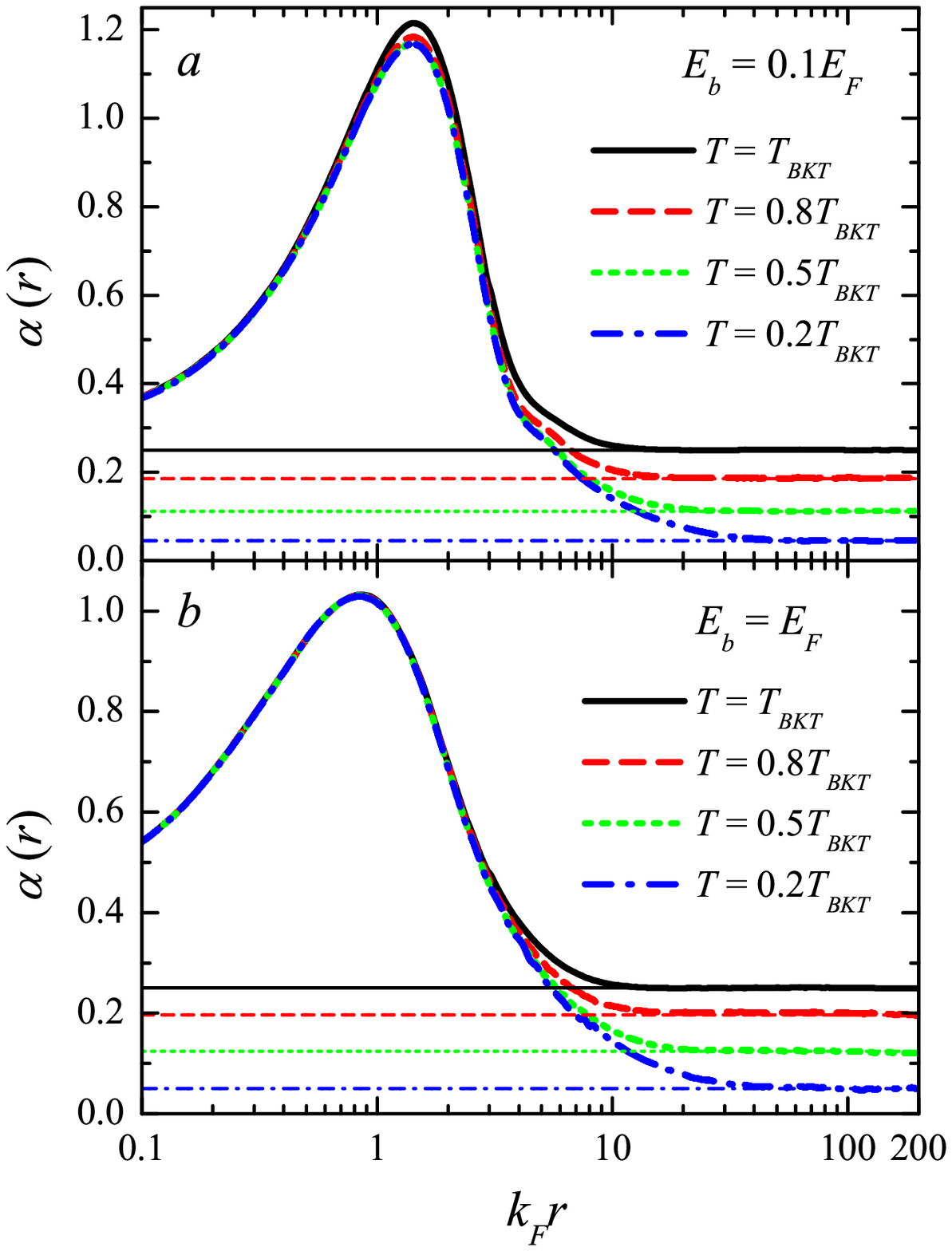}%
\caption{The function $\alpha\left(  r\right)  =r~\partial G\left(  r\right)
/\partial r$ for a Fermi gas in 2D with the binding energy $E_{b}=0.1E_{F}$
(\emph{a}) and $E_{b}=E_{F}$ (\emph{b}) at different temperatures. The
parameter $\eta$ calculated according to the BKT theory is shown by dot-dashed
lines.}%
\label{Fig2}%
\end{center}
\end{figure}

In Fig. \ref{Fig2}, the parameter $\alpha\left(  r\right)  $ determined by
formula (\ref{alp}) for a Fermi gas in 2D is plotted at different temperatures
with the same values of the binding energy as in Fig. 1. For comparison, also
the parameter $\eta\left(  T\right)  $ from the microscopic BKT theory is
shown (dot-dashed lines).%

The magnitude of $\alpha\left(  r\right)  $ can be a measure of the decay rate
for the phase correlations. We can see that the phase correlation function
rapidly falls down at intermediate distances. For sufficiently large $r$, the
parameter $\alpha\left(  r\right)  $ explicitly turns to $\eta$, as expected.
This trend is an indication of long-range correlations in a two-dimensional
Fermi gas, and confirms the suggestion that the Gaussian pair formalism is
capable to adequately describe the quasicondensate phase.

The spatial dependence of the decay rate depends relatively weakly on the
binding energy, but is temperature dependent. At small and intermediate
distances, where the decay is exponential, the correlation functions depends
only very weakly on the temperature. At distances where the decay of phase
correlations is algebraic, there is a strong temperature dependence. The
crossover value $r_{q}$ where the fast decrease of the correlation function
changes into an algebraic decay can be interpreted as the characteristic
distance at which quasi long-range order is formed. When the temperature is
close to $T_{BKT}$, we can estimate $r_{q}\gtrsim1/k_{F}$. With decreasing
temperature the distance $r_{q}$ gradually rises.%

\begin{figure}
[h]
\begin{center}
\includegraphics[
height=4.3808in,
width=3.1899in
]%
{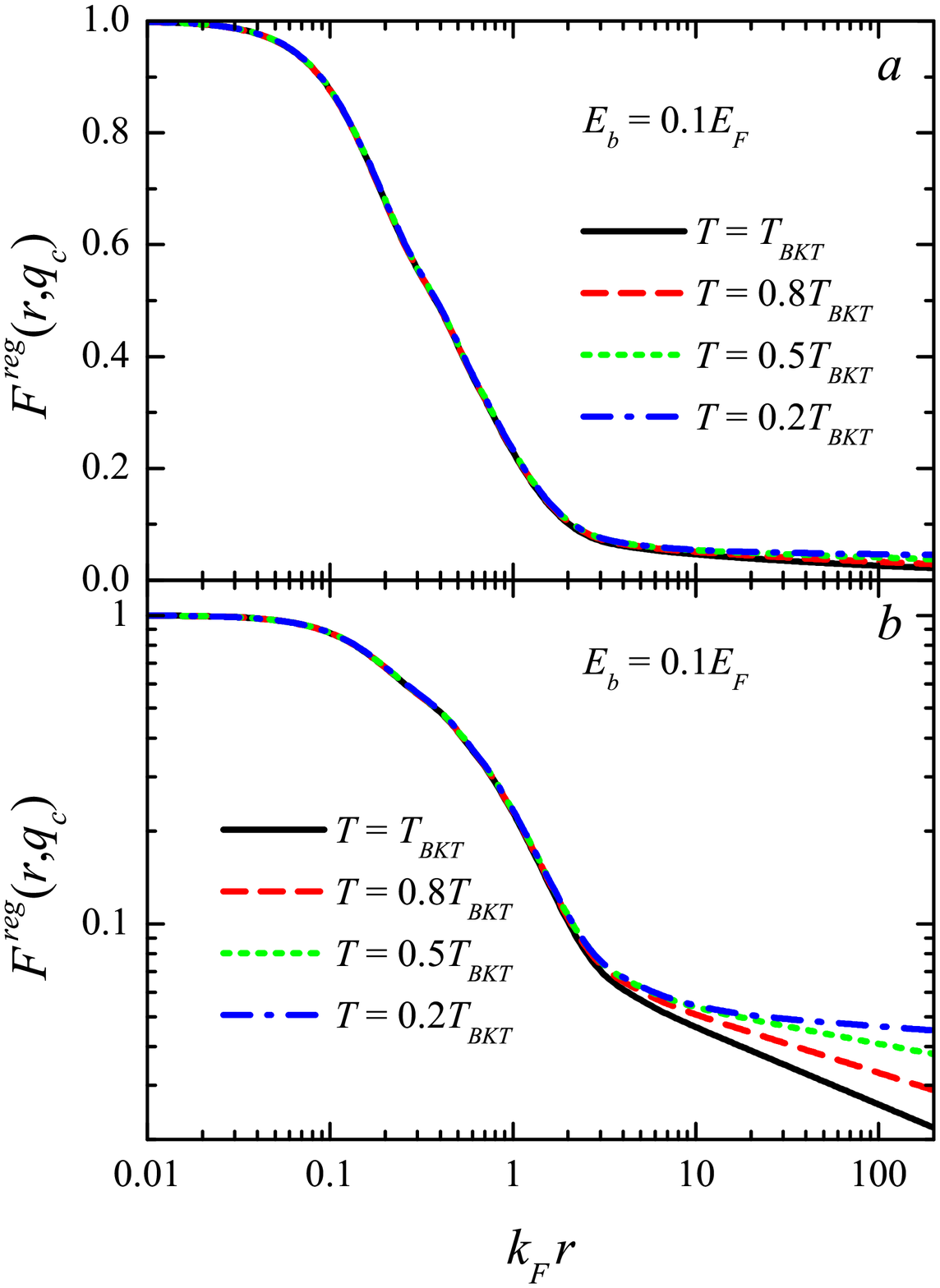}%
\caption{Cutoff-regularized phase correlation functions for the Fermi gas in
2D with the binding energy $E_{b}=0.1E_{F}$ on the linear scale (\emph{a}) and
on the logarithmic scale (\emph{b}).}%
\label{Fig3}%
\end{center}
\end{figure}

\begin{figure}
[h]
\begin{center}
\includegraphics[
height=2.7047in,
width=3.5995in
]%
{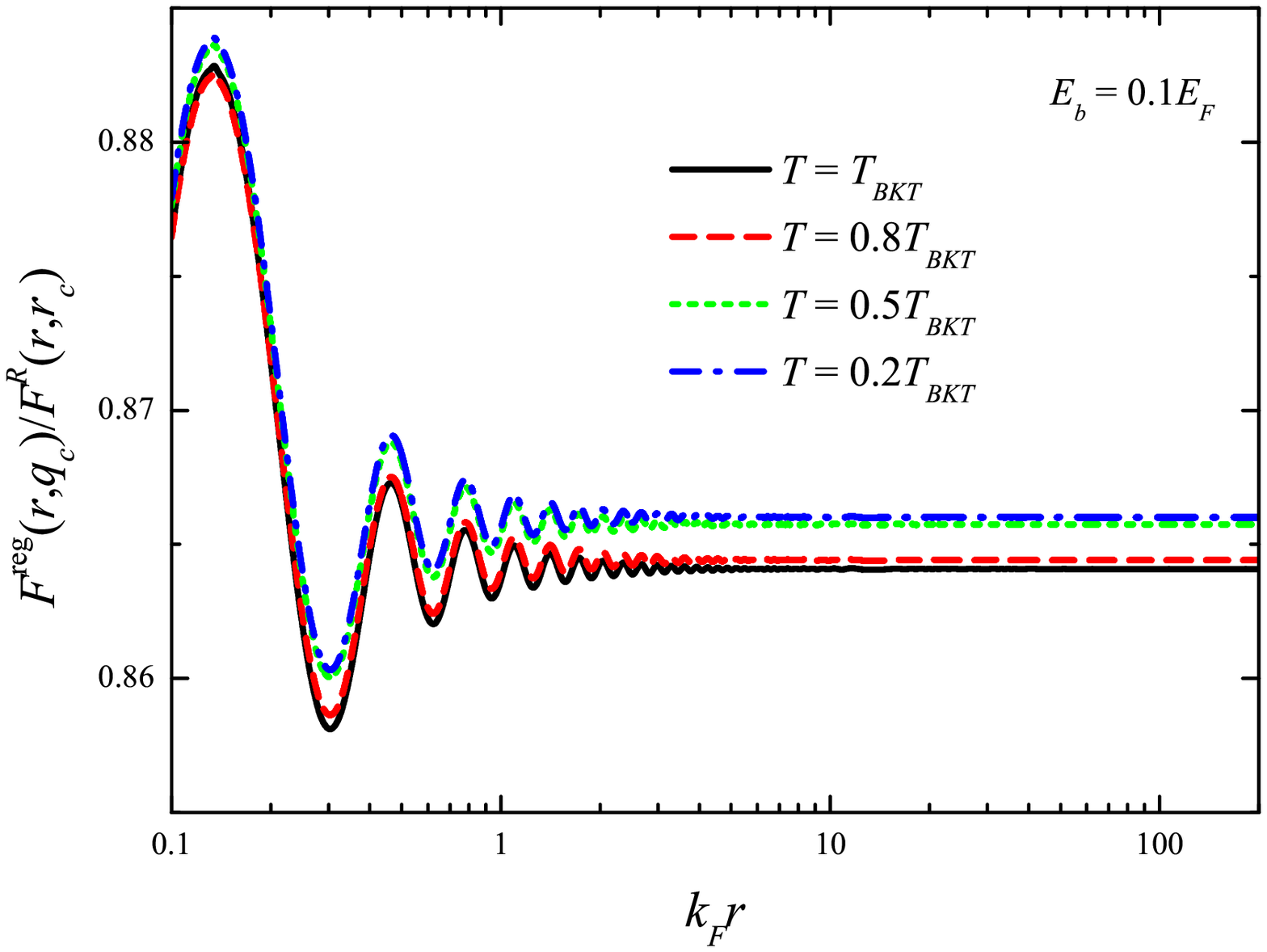}%
\caption{Ratio of the phase correlation functions $F^{reg}\left(
r,q_{c}\right)  /F^{R}\left(  r,r_{c}\right)  $ for the binding energy
$E_{b}=0.1E_{F}$ and different temperatures.}%
\label{Fig4}%
\end{center}
\end{figure}

In order to numerically verify the above conclusion that the long-range decay
of the phase correlation functions is not sensitive to a choice of the
regularization, we perform the numeric study of the regularized correlation
function $F^{reg}\left(  r,q_{c}\right)  $. Fig. \ref{Fig3} shows the phase
correlation functions $F^{reg}\left(  r,q_{c}\right)  $ obtained using the
cutoff reqularization of the ultraviolet divergence with $q_{c}=20k_{F}$ for
the binding energy $E_{b}=0.1E_{F}$ in the linear (\emph{a}) and logarithmic
(\emph{b}) scales. As distinct from the fractional correlation functions
$F^{R}\left(  r,r_{c}\right)  $, the cutoff-regularized correlation functions
contain no divergence anywhere and turn to unity at $r=0$. The coordinate
dependence of $F^{reg}\left(  r,q_{c}\right)  $ at intermediate and large
distances is very similar to that of $F^{R}\left(  r,r_{c}\right)  $: the
function $F^{reg}\left(  r,q_{c}\right)  $ decays almost exponentially at
intermediate distances, as seen from Fig. \ref{Fig3} (\emph{a}), and exhibits
an algebraic decay at rarge distances, as follows from Fig. \ref{Fig3}
(\emph{b}). To check this similarity quantitatively, we plot the ratio
$F^{reg}\left(  r,q_{c}\right)  /F^{R}\left(  r,r_{c}\right)  $ in Fig.
\ref{Fig4}. For small $r$, the regularized correlation function $F^{reg}%
\left(  r,q_{c}\right)  $ oscillates due to the factor $J_{0}\left(
q_{c}r\right)  $ at the upper bound of the integral over $q$. With increasing
$r$, these oscillations gradually fall down, vanishing at long distances, and
the ratio $F^{reg}\left(  r,q_{c}\right)  /F^{R}\left(  r,r_{c}\right)  $
tends to a constant value. The numerical check therefore confirms the
conclusion obtained above analytically: despite different regularizations, the
long-range decay is one and the same for $F^{reg}\left(  r,q_{c}\right)  $ and
$F^{R}\left(  r,r_{c}\right)  $.%

\section{Conclusions}

\label{conclusions}

We have derived the phase correlation function for an ultracold Fermi gas in
two dimensions on the basis of the Gaussian pair fluctuation action without
assuming the fluctuation field slowly varying. The resulting correlation
function describes the decay of phase correlations in the whole range of $r$,
revealing a fast decrease at intermediate distances and the characteristic
algebraic decay at large distances. This algebraic decay obtained for the
correlation functions of the phase of the fermion pair field excellently
matches the power law following from the microscopic BKT theory for Bose
gases. The appearance of the algebraic long-range order shows that the
existence of quasicondensate in two-dimensional Fermi gases can be adequately
described within the Gaussian pair fluctuation approach.

\section*{Acknowledgments}

{We are grateful to J. T. Devreese for valuable discussions. This work has
been supported by FWO-V projects G.0370.09N, G.0180.09N, G.0115.12N,
G.0119.12N, the WOG WO.033.09N (Belgium).}





\end{document}